
\magnification=1200
\hsize=15truecm
\vsize=23truecm
\baselineskip 18 truept
\voffset=-0.5 truecm
\parindent=1cm
\overfullrule=0pt

\def\Ai{\hbox{\hbox{${\cal A}$}}\kern-1.9mm{\hbox{${/}$}}}
\def\Vi{\hbox{\hbox{${\cal V}$}}\kern-1.9mm{\hbox{${/}$}}}
\def\Di{\hbox{\hbox{${\cal D}$}}\kern-1.9mm{\hbox{${/}$}}}
\def\lam{\hbox{\hbox{${\lambda}$}}\kern-1.6mm{\hbox{${/}$}}}
\def\D{\hbox{\hbox{${D}$}}\kern-1.9mm{\hbox{${/}$}}}
\def\A{\hbox{\hbox{${A}$}}\kern-1.8mm{\hbox{${/}$}}}
\def\V{\hbox{\hbox{${V}$}}\kern-1.9mm{\hbox{${/}$}}}
\def\parz{\hbox{\hbox{${\partial}$}}\kern-1.7mm{\hbox{${/}$}}}
\def\B{\hbox{\hbox{${B}$}}\kern-1.7mm{\hbox{${/}$}}}
\def\R{\hbox{\hbox{${R}$}}\kern-1.7mm{\hbox{${/}$}}}
\def\si{\hbox{\hbox{${\xi}$}}\kern-1.7mm{\hbox{${/}$}}}

\rightline{CERN.TH-6845/93}
\rightline{DFPD 93/TH/32}
\rightline{UCLA/93/TEP/13}
\rightline{NYU-TH-93/10/01}
\rightline{ENSLAPP-A442/93}
\vskip .75truecm

\centerline
{\bf BARDEEN--ANOMALY AND WESS--ZUMINO TERM IN THE}
\centerline
{\bf SUPERSYMMETRIC STANDARD MODEL}
\vskip 1.5truecm

\centerline
{\sl S. Ferrara$^{\clubsuit *}$, A. Masiero$^{\diamondsuit}$, M.
Porrati$^{\heartsuit}$ and R. Stora$^{\spadesuit}$}
\vskip .5truecm
\centerline
{\it
$^{\clubsuit}$ Theory Division, CERN, Geneva, Switzerland}
\smallskip
\centerline{\it
$^{\diamondsuit}$
Istituto Nazionale di Fisica Nucleare, Sezione di Padova, Italy}
\smallskip
\centerline
{\it
$^{\heartsuit}$ Department of Physics, New York University, New York, USA}
\smallskip
\centerline
{\it
$^{\spadesuit}$ LAPP, P.O. Box 110 F-74941, Annecy le Vieux Cedex, France}
\vskip 1truecm

\centerline{\bf Abstract}

\vskip .5truecm

\noindent
We construct the Bardeen anomaly and its related Wess--Zumino term in
the supersymmetric standard model. In particular we show that it can
be written in terms of a composite linear superfield
related to supersymmetrized
Chern--Simons forms, in very much the same way as the Green--Schwarz
term in four--dimensional string theory.
Some physical applications, such as the contribution to the g--2 of gauginos
when a heavy top is integrated out, are briefly discussed.

\vfill
\noindent
CERN TH-6845/93

\noindent
March 1993

\smallskip

\noindent
$^{*}$
Work supported in part by the Department of Energy of the United
States under contract no. DOE-ATO3-88ER40384 Task C.
\eject

\noindent
{\bf I.\ Introduction}

\vskip 0.5truecm
\noindent
The experimental fact that the mass of the top quark is much larger
than all the other quark and lepton masses has recently led several
authors to study the behaviour of the electroweak theory in the regime
of infinite top quark mass $^{[1,2]}$. This mathematical limit is not only
useful for computational purposes but it is also a rich laboratory
where one may investigate the interplay between gauge anomalies and
non--decoupling of heavy particles which are originally chiral with
respect to a gauge group ${\cal G}$ which is then spontaneously broken to an
anomaly free subgroup ${\cal H}$ by the Higgs mechanism, giving then an
${\cal H}$--invariant mass to the fermions through (${\cal G}$--invariant)
Yukawa couplings.

The best evidence of non--decoupling of heavy fermions with large
Yukawa couplings (while keeping the Higgs v.e.v. fixed) is the
phenomenon of anomaly cancellation via Wess--Zumino terms obtained
by integrating out the heavy fermions in loop--diagrams. Since ${\cal H}$
is anomaly free only the generators of ${\cal G}/{\cal H}$
are anomalous and then
the Wess--Zumino term just cancels these anomalies. In the electroweak
theory the structure group is $G=SU(2)_L\times U(1)_R$ and $H$
is the electromagnetic $U(1)$ subgroup of $G$.

In the present paper we extend this analysis to the supersymmetric
version of the standard model where now the entire (chiral) top
multiplet is integrated out and we investigate the interplay between
gauge anomalies and supersymmetric Wess--Zumino terms.

In the supersymmetric case new phenomena occur with respect to the
conventional theory $^{[2]}$:

\smallskip

\noindent
i) A gauge anomaly induces a supersymmetry anomaly, because of the
Wess--Zumino gauge--fixing condition.

\smallskip
\noindent
ii) The supersymmetric version of the Wess--Zumino term is not only
gauge variant but also violates supersymmetry. Indeed its gauge
variation cancels the gauge anomaly while its supersymmetry
transformation cancels the supersymmetry anomaly.

\smallskip

\noindent
iii) The violation of supersymmetry due to supersymmetry anomalies
also manifests itself in gauge--invariant processes involving the
Higgs particles, the vector bosons and gauginos such as the $H^\pm
W^\mp\gamma$ vertex and the magnetic moment of charged gauginos. It
also gives corrections to the Higgs scalar potential.

In the electroweak theory, due to the fact that the non--abelian
part of ${\cal G}$, based on $SU(2)$, is anomaly free, and that ${\cal H}$,
based on $U(1)_{em}$, is
abelian, the specific form of the Wess--Zumino term which cancels
the Bardeen anomaly is fairly simple and in fact can be viewed as a
four--dimensional version  of the Green--Schwarz mechanism occuring
in certain 4D strings $^{[3]}$. The main difference here is that the
antisymmetric tensor $b_{\mu\nu}$ is not a fundamental field but a composite
field made out of the Goldstone fields and of the gauge fields.

This observation allows us to derive a fairly simple supersymmetrization
of this term, working in $N=1(4D)$ superspace, without making use of the
rather complicated formulae of the non--abelian gauge supersymmetry
anomalies  derived by different groups some years ago $^{[4,5,6]}$.

\vskip 0.5truecm
\noindent
{\bf II.\ Bardeen Anomaly Revisited and its Wess--Zumino Term}

\vskip 0.5truecm
\noindent
We start by reminding some properties of the four--dimensional gauge
anomaly.
Given a gauge group ${\cal G}$ based on a structure group $G$, for each
symmetric invariant polynomial $P$ of degree 3 on Lie $G$, one can construct a
corresponding gauge anomaly:
$$
G (\alpha,A) = \int_{M_4} P(\alpha d(AdA +{1\over 2} A^3)),\eqno(1)
$$

\noindent
where $M_4$ is spacetime, $\alpha$ the gauge parameter, and $A$ the gauge
field. Here a form language has been used ($AB=A\wedge B$).
The corresponding Wess--Zumino action is $^{[7]}$

$$
\Gamma_{WZ}(A,g)=\int_{M_4}\int^1_0 dt G(g^{-1}(t)
{d\over dt} g(t), A_{g(t)} = g(t)^{-1} dg(t)+g^{-1}(t)
A g(t)),\eqno(2)
$$

\noindent
where $g(0)=identity$, $g(1)=g$ (e.g. $g(t)=e^{it\xi}$, if $g=e^{i\xi}$,
$\xi\in$ Lie $G$).

To discuss the anomalies and related W--Z action for the standard model
we have actually to depart from the consistent chiral anomaly and consider the
Bardeen anomaly $^{[8]}$ since we are interested in the electromagnetic
gauge invariant W--Z action related to the coset $(SU(2)_L\times U(1)_R/
U(1)_{em})$ rather than to a chiral group ${\cal G}$.

Here we are using at first a general method, described in ref.~$^{[7]}$,
to obtain a (generalized) Bardeen anomaly and Bardeen W--Z
action on  ${\cal G}/{\cal H}$, where ${\cal H}$ is required to be an anomaly
free subgroup
of ${\cal G}$ which means that the invariant polynomial
$P$ vanishes when Lie $G$ is restricted to Lie $H$.

The Bardeen anomaly is then

$$
G^B(\alpha,A) = G(\alpha,A) + \delta_\alpha \Delta^B (A),\eqno(3)
$$

\noindent
where

$$
G^B(\alpha,A) = 0\eqno(4)
$$

\noindent
for $\alpha\in$ Lie ${\cal H}$.

Explicit expressions are known for $G^B$, $\Delta^B$ as well as for
the W--Z action $^{[4,9]}$ which, by virtue of (3), is simply given
by

$$
\Gamma^B_{WZ} (g,A) = \Gamma^A_{WZ} (g,A) - \Delta(A) + \Delta (A_g)
\eqno(5)
$$

The point is that $\Gamma^B_{WZ}$ is ${\cal H}$ gauge invariant and
therefore depends only on the coset elements $\hat g$ in
${\cal G}/{\cal H}$.

\noindent
Hence

$$
\Gamma^B_{WZ} (g,A) = \Gamma^B_{WZ} (\hat g, A)\ .\eqno(6)
$$

In the case of the standard model a major simplification occurs because
of the particular structure of the ${\cal G}=SU(2)\times U(1)$ and $U(1)_{em}$
groups. In fact ${\cal G}$ does not have any non--abelian anomaly and
${\cal H}$ is abelian.

The invariant polynomial is taken to be

$$
T^3_0 - T_0 T^2_a\quad
(T_0 \in {\rm Lie}\; U(1), T_a\in {\rm Lie}\; SU(2)),\;\;\; a=1,2,3,\eqno(7)
$$

\noindent
and

$$
G(\alpha_0,\vec\alpha) = \alpha_0 (Tr\ F (W) F (W) - F(B) F(B)),
\;\;\; \alpha=(\alpha_0,\vec{\alpha}).
\eqno(8)
$$
The fields $W$ and $B$ correspond, respectively, to the $SU(2)$ and $U(1)$
components of the gauge connection $A$.
In this case it is easy to get $\Delta^B$ and we obtain:

$$
\Delta^B=BQ(W) + BW_3 dB,\eqno(9)
$$

\noindent
where $Q(W)$ is the Chern-Simons three-form with the propery
$$
dQ(W)=Tr\, F(W)F(W), \;\;\; (F(W)=dW+ W^2). \eqno(10)
$$

\noindent
$Q(W)$ can be written in terms of a one-parameter family of (interpolating)
gauge potentials $W(t)$ ($0\leq t\leq 1$, $W(0)=0$, $W(1)=W$) as follows
$$
Q(W)=2\int_0^1 dt Tr\, (\dot{W}(t)F(t)). \eqno(11)
$$

\noindent
For different choices of $W(t)$ we obtain gauge-equivalent $Q(W)$, i.e.
$Q(W)=Q(W') + dP$, where $P$ is a two form. The choice $W(t)=tW$ yields the
standard expression
$$
Q(W)  = Tr (WdW + {2\over 3} W^3), \eqno(12)
$$

\noindent
with gauge transformation
$$
\delta_{\vec{\alpha}}Q(W)=d (\vec{\alpha}d\vec{W}).
$$

\noindent
We work in the doublet representation of $SU(2)$ so that
$W_3 d B \rightarrow 2Tr\ W d B$ where $B=B\sigma_3/2$.

\noindent
The $U(1)$ anomaly is:

$$
G+\delta_\alpha \Delta^B = - \alpha_0 F(B) F (B) - \alpha_0
dW_3 dB\eqno(13)
$$

\noindent
The $SU(2)$ anomaly is $(\vec\alpha=\alpha_3, \alpha_i, i=1,2)$:

$$\eqalign{
\delta_{\vec\alpha} \Delta^B  & = Bd(\vec\alpha d \vec W) + B
D_{\vec\alpha} W_3 dB\cr
& = \alpha_i d W_i dB + \alpha_3 F(B) F(B) -
\epsilon_{3ij}\alpha_i W_j BdB} \eqno(14)
$$

\noindent
so that

$$\eqalign{
G^B(\alpha_i, \alpha_3-\alpha_0) & = (\alpha_3-\alpha_0)
(dB\ dB + dW_3\ dB)+\cr
& + \alpha_i dW_i dB - \epsilon_{3ij}\alpha_i W_j B\ dB}
\eqno(15)
$$

\noindent
The associated $\Gamma^B_{WZ}$ is then given by

$$\eqalign{
\Gamma^B_{WZ} (g,A) & = \Gamma^B_{WZ} (g,\vec W, B) =\cr
=\int\Bigl[ g_0 (Tr\ & F(\vec W) F (\vec W) - F(B) F(B)) -
\Delta^B (\vec W, B) + \Delta^B(\vec W_{\vec g}, B_{g_0})\Bigr]
\cr}\eqno(16)
$$

\noindent
where $(g_0, g_i)$  denotes the group element of ${\cal G}$ transforming as
$g_0\rightarrow g_0-\alpha$, $g\rightarrow \ell g$
under ${\cal G}$, and $\Delta^B$ is given by eq. (9).

{}~From eq. (16) we get

$$\eqalign{
\int\Bigl[ g_0 \Bigl( Tr F(\vec W) F(\vec W) - F(B) F(B)\Bigr) +&
\int \Bigl[ (B+dg_0) Q (\vec W_g) + \cr
+ (B+dg_0) W^g_3 dB\Bigr] - \Delta^B (W,B)  & = \int B [J(g) -
J(g=1)],\cr}\eqno(17)
$$

\noindent
where

$$
J(g,A)=Q(\vec W_g)- dg_0 dB + W^g_3 dB - dg_0 dW^g_3.\eqno(18)
$$

\noindent
Hence we recognize that

$$
\int BJ(g,A) = \Gamma_{WZ} (g,A) + \Delta^B(A_g)\eqno(19)
$$

\noindent
while

$$
-\int BJ (g=1) = - \Delta^B(A).\eqno(20)
$$

Also $J(g,A)$ can be made completely $U(1)\times SU(2)$ invariant
by noticing that we may covariantize $dg_0\rightarrow dg_0+B$ by
terms which vanish when inserted in eq. (19):

$$
J^{GW}(g,A) = Q (\vec W_g) - (dg_0 + B) dB + W^g_3 dB -
(dg_0 + B) dW^g_3\eqno(21)
$$

\noindent
which satisfies

$$
dJ^{GW} (g,A) = Tr (F(\vec W)F(\vec W)) - dB dB\eqno(22)
$$

Also the gauge invariance of eq. (18) implies that we may restrict $g$
to the coset element $U$

$$
U = e^{i\xi_i\sigma_i/2}\eqno(23)
$$

\noindent
so that

$$
J^{GW} (g,A)= J^{GW} (U,A) =  Q(W^U) -BdB - 2d(Tr W^U B)
\eqno(24)
$$

\noindent
Therefore we can finally write

$$
\Gamma^B_{WZ} (U,A) = \int B[Q(W_U) + Q(W^U_3,B) -Q (W) - Q
(W_3, B)],\eqno(25)
$$

\noindent
where $Q(W_3,B)=W_3dB -BdW_3=-d(W_3B)$.
The $U$--dependent part reproduces the consistent anomaly,
while the other part reproduces $-\delta\Delta^B$.

We may now make some further useful remarks. Let us first consider
the object ($J^{GW}$ is a three--form dual to the Goldstone--Wilczek
current):

$$
J^{GW} (U,A) - J^{GW} (1,A) = J(U,A)\quad (A=W,B).\eqno(26)
$$

\noindent
It identically satisfies

$$
dJ (U,A) =0\quad (\partial^\mu J_\mu =0).\eqno(27)
$$

\noindent
This implies that $J(U,A)$ can be locally written as the (exterior)
derivative of a two--form $T$ (antisymmetric tensor $T_{\mu\nu}$) as
follows

$$
J(U,A)= dT(U,A)\quad
(J_\mu(U,A) = \epsilon_{\mu\nu\rho\sigma}\partial_\nu T_{\rho\sigma}
(U,A))\eqno(28)
$$

\noindent
and

$$
T(U,A) =T(U,0) + \Delta T (U,A)\eqno(29)
$$

\noindent
By explicit calculation it turns out that

$$
Q(W_U) - Q(W) = - {1\over 3} (U^{-1} dU)^3 -d (dUU^{-1} W)
\eqno(30)
$$

\noindent
Hence

$$
Q(W_U) - Q(W) = d T^W (U,W)\eqno(31)
$$

\noindent
and

$$
T^W (U,W) = T^W(U,0) + \Delta T^W (U,W)\eqno(32)
$$

\noindent
with

$$
dT^W(U,0)  = - {1\over 3} Tr (U^{-1} dU)^3,\eqno(33)
$$

$$
d\Delta T^W(U,W) = - d Tr (d UU^{-1}W).\eqno(34)
$$

\noindent
This implies

$$
\Delta T^W (U,W) = - Tr d UU^{-1} W\eqno(35)
$$

$$\eqalign{
T^W(U,0) & = - Tr \int^1_0 dt U^{-1}(t) {d\over dt} U(t) U^{-1}(t)
d U(t) U^{-1}(t) d U(t)
\cr
(U(t) = & e^{i t\sigma_i/2})\cr}\eqno(36)
$$

\smallskip

$$\eqalign{
 & T^3(U,W) = - W^U_3 B + W_3 B= (W_3-W^U_3) B\cr
 & (T^3(U,W,B=0) =0).\cr}\eqno(37)
$$

\noindent
Therefore

$$\eqalign{
T(U,A) = & T^W + T^3 = T^W(U,0) - Tr dUU^{-1} W +\cr
& + (W_3 - W^U_3) B\cr} \eqno(38)
$$

\noindent
By virtue of eqs. (15), (16) and (38), the full W--Z term can
be written as

$$
\Gamma^B_{WZ} (U,A) = \int dBT(U,A)\eqno(39)
$$

\noindent
Note that the $SU(2)$ gauge transformation

$$
\delta(Q(W_U) - Q(W)) = - d\ Tr (\alpha\cdot dW)\eqno(40)
$$

\noindent
implies that

$$
\delta T^W (U,W) = - Tr\ \alpha\cdot dW.\eqno(41)
$$

\noindent
In an analogous manner, adding the $U(1)$ hypercharge transformation we have

$$
\delta(Q(W_U)-Q(W))=d(\alpha_0 dW^U_3)-d(\alpha\cdot dW),\eqno(42)
$$

\noindent
which implies:

$$
\delta_\alpha T^W = + \alpha_0 dW^U_3-\alpha\cdot dW.\eqno(43)
$$

\noindent
For $T^3$ we have:

$$\eqalign{
\delta T^3 = & - \alpha_3 dB - \alpha_0 d W^U_3 + \alpha_0
dW_3\cr
& + \alpha_0 dB + \epsilon_{3ij}\alpha_i W_j B,\cr}\eqno(44)
$$

\noindent
so that

$$\eqalign{
\delta_{\alpha_0,\vec\alpha} (T^W + T^3) & = - \alpha^T dW^T +
(\alpha_0-\alpha_3) dB +\cr
& + (\alpha_0 - \alpha_3) dW_3 + \epsilon_{3ij}\alpha_iW_j B.\cr}
\eqno(45)
$$

\noindent
We recognize that the transformation laws for $T$ are those
of an antisymmetric tensor in string theory ${[3]}$. The only difference
is that in the standard model $T$ is a composite field constructed
out of the Goldstone matrix $U$ and the gauge fields $A(W,B)$.

\vskip 0.5truecm
\noindent
{\bf III.\ Supersymmetric Gauge Anomalies in the Wess--Zumino
Gauge}

\vskip 0.5truecm
\noindent
In a supersymmetric gauge theory in D=4 dimension the supersymmetric
form of the Yang--Mills anomaly and the corresponding Wess--Zumino
lagrangian were found in refs. [4-6]. It was also found that there is
no non--trivial supersymmetric global anomaly
$^{[10]}$ in superspace (where the supersymmetry transformation laws
are linear, and Lorentz invariance is not broken).

For the physical applications supersymmetry is usually formulated in
the Wess--Zumino gauge $^{[11]}$ in which the supersymmetry algebra is
modified and the supersymmetry transformations become non--linear
$^{[12]}$.
In fact:

$$
\delta^{WZ}_{\epsilon '} = \delta_\epsilon + \delta_\Lambda\eqno(46)
$$

\noindent
where $\delta_\Lambda$ is a superspace transformation with chiral
parameters $^{[13]}$

$$
\Lambda^a = (0, {i\over\sqrt{2}}\sigma^\mu A^a_\mu
\bar\epsilon,\bar\epsilon\bar\lambda^a).
\eqno(47)
$$

\noindent
Due to this fact it follows that the superspace chiral gauge
anomaly induces, in component fields, both an ordinary gauge anomaly
$G(\alpha)$ and a (global) supersymmetry anomaly $S(\epsilon)$. If we
take

$$
\Lambda^a = ({\alpha\over 2}, {i\over\sqrt{2}}
\sigma^\mu A^a_\mu\bar\epsilon,\bar\epsilon
\bar\lambda^a)
=\Lambda^a (\epsilon,\alpha,
V_{WZ}),
\eqno(48)
$$

\noindent
and we insert this expression into the superfield formula for the
gauge anomaly with the gauge field $V$ in the Wess--Zumino gauge:

$$
G(\alpha)+S(\epsilon) = {\cal A}
(\Lambda=\Lambda(\epsilon,\alpha, V_{WZ})),\eqno(49)
$$

\noindent
eq. (49) gives for $G(\alpha)$ and $S(\epsilon)$ the same expressions
found for the solutions of the component form of the W--Z consistency
conditions $^{[14]}$:

$$
\delta_\alpha G(\alpha') - \delta_{\alpha'} G(\alpha) = G(\alpha\wedge
\alpha'),\eqno(50)
$$

$$
\delta_\epsilon G(\alpha) - \delta_\alpha S(\epsilon) = 0,\eqno(51)
$$

$$
\delta_\epsilon S(\eta) - \delta_\eta S(\epsilon) =
G(\alpha=-2i(\epsilon\sigma^m\bar\eta-\eta
\sigma^m\bar\epsilon)A_m),\eqno(52)
$$

\noindent
up to $\delta{\cal L}$ where ${\cal L}$ is a local functional
of $V_{WZ}$.

It has been shown in ref. [6] that $S(\epsilon)$ is cohomologically
non trivial in the sense that $S(\epsilon)\not = \delta_\epsilon
{\cal L}$, in the Wess--Zumino gauge.

As noted in ref. [15], the non--vanishing of $S(\epsilon)$ implies
that, in a gauge theory with anomalous fermion content, supersymmetry
is broken by loop effects even if the original classical action is
supersymmetric. A typical example is the breaking of the supersymmetric
sum rules $^{[16]}$ relating the magnetic transitions of members of charged
vector multiplets (containing the $W$ bosons) in the supersymmetric
standard model. The feedback of this phenomenon is that the integration
over the top supermultiplet produces an effective action with supersymmetry
violating terms, due to the supersymmetric Wess--Zumino term previously
discussed.

Let us make an excursion into the W--Z terms $^{[17]}$
and their supersymmetric
extension.

Because of eq. (49), the W--Z terms of a supersymmetric
theory should be given by the usual SUSY W--Z terms, derived in ref.
$^{[4]}$, computed in the W--Z gauge.

The supersymmetric W--Z term for the chiral anomaly, ${\cal A}(
\Lambda,\Lambda^\dagger,V)$, in superspace is obtained through the same
rule as the usual W--Z, namely $^{[14]}$:

$$
\Gamma^S_{WZ}(\xi,\bar\xi,V) = - \int^1_0 dt\ {\cal A}
(\Sigma^{-1}(t) \dot{\Sigma}(t),
(\Sigma(t)^{-1} \dot{\Sigma}(t))^\dagger,
\bar\Sigma(t) e^V\Sigma(t)),\eqno(53)
$$

\noindent
where $\dot{\Sigma}\equiv d\Sigma/dt$, and
$\Sigma(t) = e^{it\xi}$ is a (one--parameter family) gauge (chiral)
group element.

Since $\Gamma^S_{WZ}$ is invariant under (linear) supersymmetry
transformations, it exactly follows that:

$$
\delta_{\epsilon '} \Gamma^S_{WZ} = - {\cal A}
(\Lambda=(0,{i\over\sqrt{2}} \sigma^\mu A^a_\mu
\bar\epsilon,\bar\epsilon\bar\lambda^a))
\eqno(54)
$$

\noindent
because of eqs. (46) and (47).

Using the expression of the consistent anomaly in eq. (1), one obtains
the linearized part of the W--Z action

$$
\Gamma_{WZ}=Str(T^a T^b T^c)\int \xi^a \partial_\mu
(A^b_\nu\partial_\rho A^c_\sigma)
\epsilon^{\mu\nu\rho\sigma} dx,\eqno(55)
$$

\noindent
which has an obvious supersymmetric extension

$$
Re \;
Str (T^a T^b T^c)\int dx^4  d^2\theta i \xi^a {\cal F}^b\ {\cal F}^c
,\eqno(56)
$$

\noindent
where ${\cal F}^a_\alpha=\bar D^2 D_\alpha V^a$ is the abelian gauge field
strenght of the (non--abelian) field $V^a$.

The component expression of eq. (56) corresponds to the formula of
ref. [13]:

$$
Re\ \int dx^4 d^2\theta\ f^{bc} {\cal F}^b {\cal F}^c\eqno(57)
$$

\noindent
with

$$
f^{bc} = i\ Str (T^a T^b T^c) \xi^a\eqno(58)
$$

\noindent
Eq. (56) reduces to the exact form of the anomaly or of the W--Z
term in the abelian case with gauge transformation

$$
\delta\xi = \Lambda\eqno(59)
$$

\vskip .5truecm
\noindent
{\bf IV.\ Supersymmetric Bardeen Anomaly and its Wess--Zumino Term}

\vskip 0.5truecm
\noindent
Eq. (5) can be supersymmetrized, provided we know the
supersymmetric form of $\Delta(A)$. Then in superspace we would get

$$
\Gamma^{SB}_{WZ} (\hat\xi, V) = \Gamma^S_{WZ}(\hat\xi,V)-
\Delta^S(V)+\Delta^S(V_\xi),\eqno(60)
$$

\noindent
where

$$
V_\xi = e^{i\hat\xi^\dagger} e^V
e^{-i\hat\xi}\eqno(61)
$$

\noindent
The advantage of eq. (25) is that it involves only Chern--Simons
(C--S) forms and therefore it can be supersymmetrized in a simple
way using the C--S multiplets $^{[18]}$ $\Omega(V)$ with the property

$$
D^2\Omega = \bar D^2\Omega = Tr {\cal F}^\alpha (V)
{\cal F}_\alpha (V)\eqno(62)
$$

It is possible to give an explicit form for $\Omega(V)$ in terms of a
one-parameter family of interpolating vector superfields $V(t)$
($0\leq t \leq 1$, $V(0)=0$, $V(1)=V$), which generalizes eq. (14).
This expression reduces to the formula given in ref.~[18] for the particular
choice $V(t)=tV$.

Let us introduce the quantities
$$\eqalign{
\nabla_\alpha(t) & = e^{-V(t)} D_\alpha e^{V(t)}= D_\alpha + \phi_\alpha(t),\cr
\bar{\nabla}^{\dot{\alpha}(t)} & = e^{V(t)} \bar{D}^{\dot{\alpha}}e^{-V(t)}=
\bar{D}^{\dot{\alpha}} + \bar{\phi}^{\dot{\alpha}}(t),}\eqno(63)
$$

$$
{\cal F}_\alpha(t)  = \bar{D}^2\phi_\alpha(t),\;\;\; \bar{\cal
F}^{\dot{\alpha}}(t)=
-D^2 \bar{\phi}^{\dot{\alpha}}(t)\eqno(64)
$$

$$
H(t)= e^{-V(t)} {d\over dt} e^{V(t)},\;\;\; \bar{H}(t)=e^{V(t)}H(t)e^{-V(t)}.
\eqno(65)
$$

\noindent
By using the following property of spinorial derivatives
$$
D^\alpha \bar{D}^2 D_\alpha = \bar{D}_{\dot{\alpha}}D^2
\bar{D}^{\dot{\alpha}}, \eqno(66)
$$

\noindent
one can derive the following superspace Bianchi identity
$$
D^\alpha (e^{V(t)}{\cal F}_\alpha(t)e^{-V(t)})= e^{V(t)}\bar{D}_{\dot{\alpha}}
(e^{-V(t)}\bar{\cal F}^{\dot{\alpha}}(t) e^{V(t)})e^{-V(t)}, \eqno(67)
$$

\noindent
which implies
$$
Tr\, H(t)\{ \nabla^\alpha(t),{\cal F}_\alpha (t)\} = Tr\, \bar{H}(t)\{
\bar{\nabla}_{\dot{\alpha}}(t),\bar{\cal F}^{\dot{\alpha}} (t)\}. \eqno(68)
$$

\noindent
Here we used the property
$$
D^\alpha (e^{V(t)}{\cal F}_\alpha (t) e^{-V(t)}) = e^{V(t)} \{
\nabla^\alpha(t), {\cal F}_\alpha (t)\} e^{-V(t)}. \eqno(69)
$$

\noindent
By means of eq.~(68) and using the same argumets as in ref.~[18] one easily
finds
$$\eqalign{
\Omega(V) = & 2\int^1_0 dt\ Tr \Bigl\{ [\nabla^\alpha(t),H(t)]
{\cal F}_\alpha (t) +\cr
+ & [\bar\nabla_{\dot\alpha}(t),\bar{H}(t)] \bar{\cal F}^{\dot\alpha} (t) +
H(t)\{\nabla^\alpha, {\cal F}_\alpha(t)\} \Bigr\}\cr}\eqno(70)
$$

\noindent
Notice that, in analogy with the bosonic case, the choice of the interpolating
path $V(t)$ introduces an arbitrariness in the definition of $\Omega$. Two
different C-S multiplets, $\Omega_1(V)$ and $\Omega_2(V)$, corresponding to
two different paths $V_1(t)$, $V_2(t)$, obey, by virtue of eq.~(62)
$$
\bar{D}^2 [\Omega_1(V)-\Omega_2(V)]=D^2 [\Omega_1(V)-\Omega_2(V)]=0.
\eqno(71)
$$

\noindent
This equation implies that
$$
\Omega_1(V)=\Omega_2(V)+L(V), \eqno(72)
$$

\noindent
where $L(V)$ is a linear multiplet $^{[11]}$.
By dimensional arguments it has the form
$$
L(V)=D^\alpha \bar{D}^2P_\alpha(V) + h.c., \eqno(73)
$$

\noindent
where $P_\alpha (V)$ is an unconstrained superfield built in terms of $V$ and
$D_\alpha V$.

Eqs.~(71),(72) express the gauge equivalence of $\Omega_1$ and $\Omega_2$ with
respect to superspace gauge transformations.

In order to supersymmetrize the counterterm $\Delta^B$ of eq.~(9)
we also need the analog of the $Q(W_3,B)$ mixed C-S three-form. It reads
$$
\Omega(\hat{W}_3,B) = D^\alpha \hat{W}_3 {\cal F}_\alpha(B) +
\bar{D}_{\dot\alpha}
\hat{W}_3\bar{\cal F}^{\dot\alpha} (B) + \hat{W}_3D^\alpha
{\cal F}_\alpha (B)- (\hat{W}_3 \leftrightarrow B).\eqno(74)
$$

\noindent
The superfield $\hat{W}_3$ is defined below, by eqs.~(82),(83).
Note that $\Omega(\hat{W}_3,B)$ is a linear multiplet since
$D^2 \Omega(\hat{W}_3,B) = \bar{D}^2 \Omega(\hat{W}_3,B) =0 $.

The supersymmetrized form of $\Delta^S$ gives rise to the supersymmetric
Bardeen anomaly provided that under the infinitesimal $\Lambda_3$ supergauge
transformation
$$
\delta e^{\sigma_a W_a/2}= -{i\over 2}\sigma_3 \Lambda_3^\dagger
e^{\sigma_a W_a/2} +{i\over 2}e^{\sigma_a W_a/2}\sigma_3\Lambda_3^\dagger,
\eqno(75)
$$

\noindent
$\hat{W}_3$ and $\Omega(W)$ transform as follows
$$
\delta_{\Lambda_3} \hat{W}_3 = i(\Lambda_3 -\Lambda^\dagger_3),
\eqno(76)
$$

$$
\delta_{\Lambda_3} \Omega(W) = {i\over 2}D^\alpha
\bar{D}^2(\Lambda_3D_\alpha \hat{W}_3) + h.c.. \eqno(77)
$$

\noindent
If this is the case, the supersymmetric Bardeen anomaly takes the form
$(i=1,2)$
$$
\eqalign{
{\cal A}^B(W,B,\Lambda_0-\Lambda_3,\Lambda_i)=&
-{i\over 2}\int d^2\theta (\Lambda_0-\Lambda_3)
{\cal F}^\alpha(B){\cal F}_\alpha (B) \cr
& -{i\over 2}\int d^2\theta (\Lambda_0-\Lambda_3){\cal F}^\alpha(\hat{W})
{\cal F}_\alpha (B) \cr
& + h.c. +\delta_{\Lambda_i}\Delta^S.}
\eqno(78)
$$

\noindent
This equation is obtained by adding to the $SU(2)$-invariant mixed anomaly
$$
i\int d^2\theta \Lambda_0[ Tr\, {\cal F}^\alpha(W)  {\cal F}_\alpha(W)
-  {1\over 2}{\cal F}^\alpha(B)  {\cal F}_\alpha(B)], \eqno(79)
$$

\noindent
the $SU(2)\times U(1)$ gauge variation of $\Delta^S$, given by
$$
\Delta^S= -\int d^4\theta B(\Omega(W) +{1\over 2}\Omega(\hat{W}_3,B)).
\eqno(80)
$$

\noindent
It is worth commenting on the uniqueness of ${\cal A}^B$,
given by eq.~(78), and its
related Wess--Zumino term $\Gamma^{SB}_{WZ}$, given by eq.~(60). ${\cal A}^B$
is defined only up to an electromagnetic gauge-invariant counterterm
$\Delta^I$
$$
\delta_{e.m.}\Delta^I=0.
$$

\noindent
In the bosonic case, the expression given in eq.~(15) is unique if CP
conservation is assumed$^{[1]}$.
In the supersymmetric case, due to the non-polynomial structure of the
gauge-field transformations, this result does not follow in any obvious way
from the non-supersymmetric case.

To obtain eqs.~(76) and~(77) we first have to construct new gauge field
variables $\hat{W}_3$, $\hat{W}_i$ such that, under $\sigma_3\Lambda_3$ gauge
transformations
$$
\eqalign{ \delta \hat{W}_3 =& i (\Lambda_3 -\Lambda_3^\dagger), \cr
\delta \hat{W}_i =& -{1\over 2}
\epsilon_{ij}(\Lambda_3 +\Lambda_3^\dagger)\hat{W}_j,
\;\;\; (\epsilon_{12}=-\epsilon_{21}=1).} \eqno(81)
$$

\noindent
As explained in the appendix, this is achieved by defining
$$
\hat{W}_i= {\sinh \xi\over 2\xi} W_i, \eqno(82)
$$

$$
e^{\hat{W}_3/2}(1+\hat{W}^2_i)^{1/2}=\cosh \xi + {\sinh \xi\over 2\xi}
W_3,\;\;\; \xi^2={W^2\over 4}={W_aW_a\over 4}.
\eqno(83)
$$

\noindent
In terms of these variables we have
$$
e^V=e^{\sigma_aW_a/2}=e^{\sigma_3W_3/2}(1+\hat{W_i}^2)^{1/2} +
\hat{W}_i\sigma_i . \eqno(84)
$$
To obtain eq.~(77) we first define an $\hat{\Omega}(W)$ through a particular
interpolating path $W(t)$, such that, under a $\sigma_3\Lambda_3$
transformation
$$
e^{W(t)}\rightarrow e^{-i\Lambda^\dagger_3(t)\sigma_3/2} e^{W(t)}
e^{i\Lambda_3(t)\sigma_3/2}, \eqno(85)
$$

\noindent
for some specific choice of the gauge-transformation
interpolating path $\Lambda_3(t)$.

If these interpolating paths exist then
$$
\delta_{\Lambda_3}\hat{\Omega}(W)=2i\int_0^1dt D^\alpha(Tr\,
\dot{\Lambda} {\cal F}_\alpha (t)) + h.c.. \eqno(86)
$$

\noindent
Here $\dot{\Lambda}(t)\equiv (\sigma_3 d\Lambda_3(t)/dt)/2$.

One easily verifies that eq.~(85) holds when the path is chosen as follows
$$
\eqalign{(\hat{W}_3(t),\hat{W}_i(t))=& (2t\hat{W}_3,0),\;\;\; 0\leq t \leq
{1\over 2}, \cr
(\hat{W}_3(t),\hat{W}_i(t))=& (\hat{W}_3,(2t-1)\hat{W}_i),\;\;\; {1\over 2}
\leq t \leq 1 .} \eqno(87)
$$

\noindent
and
$$
\Lambda_3(t)=2t\Lambda_3,\;\;\; 0\leq t \leq {1\over 2}, \;\;\;
\Lambda_3(t)=\Lambda_3,\;\;\; {1\over 2} \leq t \leq 1.
\eqno(88)
$$

\noindent
Inserting eqs.~(87),(88) into eq.~(86) we finally obtain eq.~(77).

Let us notice that at the linearized level, that is for the terms in
$\Omega$ quadratic in the gauge superfields, the path given in eq.~(87) gives
the same C-S form of the path $W(t)=tW$. Thus, eq.~(77) reduces to the abelian
gauge transformation eq.~(21). The same holds for the complete non-abelian
transformation in the Wess--Zumino gauge.

We therefore obtain a superfield formula for $\Gamma^{SB}_{WZ}$ in
the supersymmetric standard model as follows

$$\eqalign{
\Gamma^{SB}_{WZ} & (U= e^{i\xi_i\sigma_i/2}, W,B)=\cr
= \int d^4\theta & \ d^4x B[\hat{\Omega}(W_\xi) - \hat{\Omega}(W) +
{1\over 2}\Omega(\hat{W}^\xi_3,B)-{1\over 2}\Omega(\hat{W}_3,B)],}\eqno(89)
$$

\noindent
with

$$
W_\xi\rightarrow U^\dagger e^W U.\eqno(90)
$$

\noindent
At the linearized level it reduces to

$$\eqalign{
Re & \int d^2\theta (i\xi^a {\cal F}_W^a {\cal F}_B + i \xi^3 {\cal F}_B {\cal
F}_B)\cr
( {\cal F}_V & = \bar D\bar D D_\alpha V\ {\rm here})\cr}
\eqno(91)
$$

\noindent
and its component expression gives

$$
i\xi^a=(H^a, G^a, \lambda^a, h_1^a, h_2^a)
$$

$$\eqalign{
\Gamma_{WZ}^{SB} = & \int d^4z {1\over 2}\Bigl[ H^a(F^a_{\mu\nu}
F_{\mu\nu} (B) + i
\bar\lambda^a\parz\lambda_B + i\bar\lambda^B \parz \lambda^a\cr
& - 2 D^a D^B\cr
+ & G^a (-{1\over 2} F^a_{\mu\nu} \tilde F^B_{\mu\nu} - i\partial_\mu
(\bar\lambda^a \gamma_5 \gamma_\mu \lambda^B)\cr
- i & \bar\chi^a (D^a \gamma_5\lambda^B + D^B \gamma_5 \lambda^a +
 F^a_{\mu\nu}\sigma^{\mu\nu} \lambda^B+\cr
+ &  F^B_{\mu\nu} \sigma^{\mu\nu} \lambda^a)+\cr
+ & i h^a_1 \bar\lambda^a\lambda^B + i h^a_2 \bar\lambda^a \gamma_5
\lambda^B\Bigr]\cr
+ & \Bigl[ H^3 ({1\over 2} F^{B\ 2}_{\mu\nu}+ i\lambda^B\parz
\lambda^B - D^{B\ 2})\cr
+ & G^3 (-{1\over 4} F^B_{\mu\nu} \tilde F^B_{\mu\nu} - {i\over 2}
(\bar\lambda^B \gamma_5 \gamma_\mu \lambda^B))\cr
- & i\bar\chi^3 (D^B \gamma_5 \lambda^B +  F^B_{\mu\nu}
\sigma^{\mu\nu} \lambda^B)\cr
+ & {i\over 2} h^3_1 \bar\lambda^b\lambda^B + {i\over 2} h^3_2
\bar\lambda^B \gamma_5 \lambda^B\Bigr].\cr}\eqno(92)
$$

The terms proportional to $G^a$ and $G^3$ in $\Gamma_{WZ}^{SB}$
is the ordinary W--Z term whose gauge variation gives the gauge
anomaly, while the terms proportional to $\chi^a$,
$\chi^3$ and $h^a$, $h^3$
are responsible for the supersymmetric anomaly due to the anomalous
supersymmetry transformation of $\chi$ and $h$. Finally, the term
proportional to the pseudogoldstone $H^a$ does not contribute to the
anomaly, but is required by supersymmetry.

\noindent
In the supersymmetric case the multiplet

$$
L^W(U,W) = \hat{\Omega}(W^U) - \hat{\Omega} (W)\eqno(93)
$$

\noindent
is a linear multiplet $^{[19]}$ $(\bar D\bar D L^W=DDL^W=0)$ and so it is
the multiplet

$$
L^3(U,W,B)  =\Omega(\hat{W}_3^U, B) - \Omega(\hat{W}_3,B).  \eqno(94)
$$

\noindent
In fact $\Omega(\hat{W}_3,B)$ itself is a linear multiplet given by
$$
\Omega(\hat{W}_3,B)=D^\alpha \bar{D}\bar{D} (\hat{W}_3
{\mathop D\limits^{\leftrightarrow}}_\alpha B) + h.c. .
\eqno(95)
$$

\noindent
Then we conclude that

$$
L(U,W,B)=L^W (U,W) + {1\over 2}L^3 (U,W,B)\eqno(96)
$$

\noindent
is a linear multiplet with $(L(U=1)=0)$.

So the full supersymmetric W--Z term can be written as

$$
\int d^4\theta d^4x BL (U,W,B),   \eqno(97)
$$

\noindent
and it is precisely analogous to the supersymmetric form of the
Green--Schwarz term $^{[3,18]}$ with a composite linear multiplet $L(U,W,B)$.

\noindent
Eq. (97) is the supersymmetric version of eq. (25).

A linear multiplet can be locally solved as

$$
L=D^\alpha \bar D^2 U_\alpha + \bar D_{\dot\alpha} D^2
\bar U^{\dot\alpha}\ .\eqno(98)
$$

\noindent
(with $U_\alpha$ defined up to terms $U_\alpha=i D_\alpha Z (Z=Z^\dagger)$).

\noindent
At the linearized level in the $U\sim 1 +i\xi^i\sigma_i/2$

$$
U_\alpha={i\over 2}\xi^i D_\alpha W^i + {i\over 2}\xi^3D_\alpha W^3 .\eqno(99)
$$

\noindent
The most general form of $U_\alpha(U,W,B)$ can be constructed from the
identity
$$
L(U,W,B)=\int_0^1 dt {d\over dt} [\hat{\Omega}(\hat{W}^{U(t)}) +
{1\over 2}\Omega(W_3^{U(t)},B)], \eqno(100)
$$

\noindent
with $U(t)=e^{i\xi(t)}$ ($\xi(0)=0$, $\xi(1)=\xi$).
Indeed, if
$$
\delta_\Lambda (\hat{\Omega}(W)+ {1\over 2}\Omega(\hat{W}_3,B)) =
D^\alpha \bar{D}^2 \Gamma_\alpha (W,B,\Lambda) + h.c.,
\eqno(101)
$$

\noindent
then we find the following formula for $U_\alpha$
$$
U_\alpha(W,B,U)= \int_0^1dt \Gamma_\alpha(W^{U(t)},B,U^{-1}(t){d\over
dt}U(t)).
\eqno(102)
$$

By means of eq. (98), integrating by part eq. (97), we obtain the
supersymmetric counterpart of eq. (39):

$$
Re\int d^4 x d^4\theta {\cal F}^\alpha(B) U_\alpha(U, U^\dagger,V).
\eqno(103)
$$

\noindent
At the linearized level, by using eq.~(99), the above formula reproduces
eq.~(91).
\vskip 0.5truecm
\noindent
{\bf V.\ Conclusions}

\vskip 0.5truecm
\noindent

We would like finally to comment on some physical effects of the
supersymmetric Wess--Zumino term as given by eq. (97).

Already at the linearized level (eq. (92)) we see that, beyond
the usual terms of the non supersymmetric case, there is a
contribution to the $HW\gamma$ coupling and to the magnetic
moment of the gauginos. Also the auxiliary fields give a correction
to the scalar Higgs potential as well as to the gaugino Yukawa
couplings. These terms can be obtained, by explicit computation,
sending to infinity the top (or bottom) Yukawa couplings in loop
diagrams, as already shown for the g--2 in some examples $^{[15,2,20]}$.
Quite independently, Wess--Zumino terms have broader applications in any
context where some fermions contributing to anomalies are replaced by some
bosonic fields interacting with gauge fields.
\vskip .5truecm
\noindent
{\bf Acknowledgements}
\smallskip
\noindent
S. Ferrara and M. Porrati would like to thank, respectively,
the Santa Barbara Institute for Theoretical Physics and the Aspen Institute
for Physics, where part of this work was done, for their kind hospitality.

\vskip .5truecm
\noindent
{\bf Appendix}
\vskip .5truecm
\noindent
In this appendix we describe in some details how to find the hatted variables
$\hat W_a$, $a=1,2,3$.

The gauge transformation $^{[11]}$
$$
e^V \rightarrow e^{-i\Lambda^\dagger} e^V e^{i\Lambda}
\eqno{A.1}
$$
implies the following (infinitesimal) gauge transformation for the
Lie algebra valued superfield $V$ $^{[21]}$
$$
\delta V= i{\cal L}_{V/2} \Delta + {\cal L}_{V/2} \coth {\cal L}_{V/2}
\Sigma,
\eqno{A.2}
$$
with $\Delta=\Lambda +\Lambda^\dagger$, $\Sigma = i(\Lambda-\Lambda^\dagger)$.
For the group $G=SU(2)$ with
$$
\Lambda={1\over 2} \Lambda_a\sigma_a,\;\;\; V={1\over 2} W_a \sigma_a,
\eqno{A.3}
$$
we obtain the component expression
$$
\delta W_a = {1\over 2} \epsilon_{abc}\Delta_b W_c +{1\over
4}f(\xi)(W^2\delta_{ab} -W_aW_b)\Sigma_b +\Sigma_a,
\eqno{A.4}
$$
with
$$
f(\xi)=\xi^{-2}(\xi\coth\xi-1) ,\;\;\; \xi^2\equiv {W^2\over 2}.
\eqno{A.5}
$$
Notice that $\delta W^2=2W_a\Sigma_a$.

In the Wess--Zumino gauge
$$
\delta W_a=\Sigma_a + {1\over 2}\epsilon_{abc}\Delta_bW_c.
\eqno{A.6}
$$
It is easy to see that no field redefinition of the form
$$
\hat{W}_a=W_a(W),\;\;\; \forall a
\eqno{A.7}
$$
exists, such that eq.~A.6 holds in any supersymmetric gauge. However it
is possible to define new $\hat{W}_a$ such that~A.6 is verified for
$\Lambda=\Lambda_3\sigma_3/2$.
Indeed, let us set
$$
\hat{W}_i=g(\xi) W_i,\;\;\; i=1,2.
\eqno{A.8}
$$
{}From eq.~A.6 and~A.4, we obtain the following differential equation
$$
\xi f(\xi) ={g'(\xi)\over g(\xi)},
\eqno{A.9}
$$
whose solution is
$$
g(\xi)= \lambda {\sinh\xi \over \xi}.
\eqno{A.10}
$$
We set the arbitrary constant of integration $\lambda$ equal to $1/2$; thus,
$$
\hat{W}_i={\sinh\xi \over 2\xi} W_i.
\eqno{A.11}
$$
Next, we define $\hat{W}_3$ as
$$
\hat{W}_3= W_3 + F(\hat{W}^2,\xi),\;\;\;\hat W^2\equiv \hat W_1^2 + \hat
W_2^2,
\eqno{A.12}
$$
and demand
$$
\delta_{\Lambda_3}\hat W_3 =\Sigma_3.
\eqno{A.13}
$$
Note that $\delta_{\Lambda_3}\hat W^2=0$.
Eq.~A.13 gives a first-order differential equation for $F$
$$
{dF\over 2\hat W^2}= (\sinh^2\xi -\hat W^2 )^{-1/2} d\bigl({\xi \over
\sinh\xi}\bigr),
\eqno{A.14}
$$
whose solution is
$$
F= -W_3 + 2\log (\cosh\xi + {\sinh\xi \over \xi} {W_2\over 2}) + \log f(\hat
W^2).
\eqno{A.15}
$$
In this equation, $f$ is an arbitrary function, and we used the relation
$$
\sinh^2\xi -\hat W^2 = {\sinh^2\xi \over \xi^2} {W_3^2\over 4}.
\eqno{A.16}
$$
The above equations give
$$
e^{\hat W_3/2}= (\cosh\xi + {\sinh \xi \over 2\xi} W_3) f^{1/2}(\hat W^2).
\eqno{A.17}
$$
By choosing $f(\hat W^2)=(1+\hat W^2)^{-1}$ we arrive at equation~(83). This
particular form of the function $f(\hat W^2)$ has been chosen so that eq.~(84)
holds.
\vskip .5truecm
\noindent
{\bf References}

\vskip 0.5truecm
\noindent

\item{[1]} T. Sterling and M. Veltman, Nucl. Phys. \underbar{B189}
(1981) 557;
\item{} E. D'Hoker and E. Fahri, Nucl. Phys. \underbar{B248} (1984)
59; Nucl. Phys. \underbar{B248} (1984) 77;
\item{} J. Preskill, Ann. Phys. (N.Y.) \underbar{210} (1991) 323;
\item{} G.L. Lin, H. Steger and Y.P. Yao, Phys. Rev. \underbar{D44}
(1991) 2139.
\item{} F. Feruglio, L. Maiani and A. Masiero, Nucl. Phys.
\underbar{B387} (1992) 523.
\item{} E. D'Hoker, Phys. Rev. Lett. \underbar{69} (1992) 1316.

\item{[2]}
S. Ferrara, A. Masiero and M. Porrati, Phys.
Lett. \underbar{B301} (1993) 358.

\item{[3]} M. Green and J. Schwarz, Phys. Lett. \underbar{149B} (1984)
17.
\item{} M. Dine, N. Seiberg and E. Witten, Nucl. Phys. \underbar{B289} (1987)
589.
\item{[4]} G. Girardi, R. Grimm and R. Stora, Phys. Lett. \underbar{156B}
(1985) 203;
\item{} L. Bonora, P. Pasti and M. Tonin, Phys. Lett. \underbar{156B}
(1985) 341.

\item{[5]} E. Guadagnini, K. Konishi and M. Mintchev, Phys. Lett.
\underbar{157B} (1985) 37;
\item{} N.K. Nielsen, Nucl. Phys. \underbar{B244} (1984) 499.

\item{[6]} H. Itoyama, V.P. Nair and H. Ren, Nucl. Phys. \underbar{B262}
(1985) 317.
\item{} E. Guadagnini and M. Mintchev, Nucl. Phys. \underbar{B269}
(1986) 543.

\item{[7]} W. A. Bardeen and B. Zumino, Nucl. Phys. \underbar{B244}
(1984) 421.
\item{} B. Zumino, Wu Yong--Shi and A. Zee, Nucl. Phys. \underbar{B239}
(1984) 477.
\item{} J. Manez, R. Stora and B. Zumino, Comm. Math. Phys. \underbar{102}
(1985) 157.

\item{[8]} W.A. Bardeen, Phys. Rev. \underbar{184} (1969) 1848.

\item{[9]} E. Witten, Nucl. Phys. \underbar{B223} (1983) 422,433.
\item{} Chou Kuang--chao, Guo Han--ying, Wu Ke and Song Xing--chang,
Phys. Lett. \underbar{134B} (1984) 67.
\item{} H. Kawai and S.-H.H. Tye, Phys. Lett. \underbar{140B} (1984)
403.

\item{[10]} O. Piguet and K. Sibold, Nucl. Phys. \underbar{B247}
(1984) 484.
\item{} J. Dixon, Class. Quant. Grav. \underbar{7} (1990) 1511.
\item{[11]} J. Wess and B. Zumino, Phys. Lett. \underbar{B78}
(1974) 1; S. Ferrara and B. Zumino, Nucl. Phys. \underbar{B79}
(1974) 413.

\item{[12]} B. de Wit and D.Z. Freedman, Phys. Rev. \underbar{D12}
(1975) 2286

\item{[13]} E. Cremmer, S. Ferrara, L. Girardello and A. Van Proeyen,
Nucl. Phys. \underbar{B212} (1983) 413.

\item{[14]} B. Zumino in ``Geometry, Anomaly and Topology", (World
Scientific, Singapore, 1985).

\item{[15]} C.L. Bilchak, R. Gastmans and A. Van Proeyen, Nucl. Phys.
\underbar{B273} (1986) 46.

\item{[16]} S. Ferrara and M. Porrati, Phys. Lett. \underbar{288B}
(1992) 85.

\item{[17]} J. Wess and B. Zumino, Phys. Lett. \underbar{37B} (1971)
95.

\item{[18]} S. Cecotti, S. Ferrara and M. Villasante, Int. J. Mod. Phys.
\underbar{A2} (1987) 1839.

\item{[19]} S. Ferrara, J. Wess and B. Zumino, Phys. Lett.
\underbar{51B} (1974) 239.

\item{[20]} S. Ferrara and A. Masiero, preprint CERN TH-6846/93,
to appear in the Proc. of the 26th Workshop, Eloisatron Project,
``From Superstrings to Supergravity", Erice, December 1992.

\item{[21]} J. Wess and J. Bagger, ``Supersymmetry and Supergravity'', p. 46
(Princeton University Press, Princeton NJ, 1992).

\bye